# Concept Mapping as an Instrument for Evaluating an Instruction Unit on Holography

# Concept Maps als Evaluierungsinstrumente einer Unterrichtseinheit zur Holographie

Martin Erik Horn, Helmut F. Mikelskis

University of Potsdam, Physics Education Research Group,
Am Neuen Palais 10, 14469 Potsdam, Germany
E-Mail: marhorn@rz.uni-potsdam.de – mikelskis@rz.uni-potsdam.de

**Abstract**
Due to its amazing three-dimensional effects, holography is a very motivating, yet very demanding subject for physics classes at the upper level in school. For this reason an instruction unit on holography that supplement holographic experiments with computer-supported work sessions and a simulation program was developed.
The effects of the lessons on holography were determined by a pre-post-test design. In addition to videotaping the lessons, knowledge and motivational tests as well as student interviews, students were asked to prepare concept maps, which were used to track processes of model construction.
The way this knowledge was applied largely depends on the students' understanding of models. In particular it was shown that the participating students' demonstrated capacity for distinguishing between the different models of light is of great importance. Only students with a developed capacity for distinguishing between models are able to reason in an problem-oriented manner. They recognize the limits of models and are not only able to specifically designate changes from models and levels of models, but also to apply these in a physically correct manner.

**Kurzfassung**
Aufgrund ihrer verblüffenden Dreidimensionalität stellt die Holographie ein sehr motivierendes, jedoch auch anspruchsvolles Unterrichtsthema für Oberstufenkurse Physik dar. Deshalb wurde an der Universität Potsdam eine Unterrichtsreihe zur Holographie entwickelt, die Experimente zur Holographie durch computergestützte Arbeitsphasen und ein Simulationsprogramm ergänzt.
Im Rahmen der Evaluation dieser Reihe wurden die Wirkungen des Unterrichts zur Holographie mit Hilfe eines Pre-Post-Testdesigns ermittelt. Neben der Videographie des Unterrichts, einem Wissens- und Motivationstest und Schülerinterviews wurde die Anfertigung von Concept Maps durch die Schülerinnen und Schüler genutzt, um Modellbildungsprozesse zu verfolgen. Der Abgleich der Daten zeigte, dass ein höchst signifikanter Zuwachs des Wissens zur Holographie erfolgte.
Der Umgang mit diesem Wissen ist jedoch in starkem Maße abhängig vom Modellverständnis, das die Schüler aufweisen. Insbesondere konnte gezeigt werden, dass der von den Probanden gezeigten Trennschärfe zwischen den Modellen des Lichts (Strahlenmodell, Wellenmodell, klassisches und quantenoptisches Teilchenmodell sowie Zeigermodell) eine entscheidende Bedeutung zukommt. Ist keine Trennschärfe vorhanden, so springen die Schüler zur Erklärung holographischer Prozesse beliebig und physikalisch inkonsistent zwischen den Modellen hin und her. Ihr Wissen ist rein reproduzierend. Dagegen gelingt es Schülern mit ausgeprägter Trennschärfe, bewusst und problemorientiert zu argumentieren, die Modellgrenzen zu erkennen und Wechsel von Modellen und Modellebenen nicht nur ausdrücklich zu benennen, sondern diese physikalisch sinnrichtig anzuwenden und zu begründen.

| Contents | Inhalt |
|---|---|
| 1. Introduction | 1. Einleitung |
| 2. Methods | 2. Untersuchungsmethoden |
| 3. Results | 3. Ergebnisse |
| 4. Conclusions and implications | 4. Schlussfolgerungen und Konsequenzen |
| 5. Bibliography | 5. Literaturangaben |

**1. Introduction**

Due to its amazing three-dimensional effects, holography is a very motivating, yet very demanding subject for physics classes at the upper level at school. For this reason an instruction unit on holography that supplement holographic experiments with computer-supported work sessions and a simulation program was developed at the University of Potsdam. In addition to the program phenOpt [1], a simulation program on holography [3] is implemented. This unit, which consisted of 12 hours, was tested in grade 13. Besides analyzing the overall effects, the evaluation tracked the model construction processes of students and examined them with the following scientific questions:

- What is the impact of models used by the students on the comprehension of the basic subject matter of holography, and what degree of distinction exists between alternative model conceptions?

The research was based on the hypothesis that students, who have a developed capacity for distinguishing between alternative model conceptions, use models confidently and appropriately in a holographic context, while students without this capacity for distinguishing between models are not able to explain holography consistently.

**2. Methods**

The effects of the lessons on holography were determined by a pre-post-test design. In addition to videotaping the lessons, knowledge and motivational tests as well as student interviews, students were asked to prepare concept maps, which were used to track processes of model construction.

The concept maps to be worked on by the students consisted of ten concepts, which had been extrapolated from a feasibility study with university students. The participating students were asked to mark the proposition arrows directly after entering them. Modal maps of the pre-test, the post-test as well as the long-term test were generated according to [4] from the concept map data of all students, which were then analyzed based on their contents,

**1. Einleitung**

Aufgrund ihrer verblüffenden Dreidimensionalität stellt die Holographie ein sehr motivierendes, jedoch auch anspruchsvolles Unterrichtsthema für Oberstufenkurse Physik dar. Deshalb wurde an der Universität Potsdam eine Unterrichtsreihe zur Holographie entwickelt, die Experimente zur Holographie durch computergestützte Arbeitsphase und ein Simulationsprogramm ergänzt. Eingesetzt werden neben den Programmen phenOpt [1] ein Simulationsprogramm zur Holographie [3]. Erprobt wurde die Unterrichtseinheit im Umfang von 12 Stunden in der 13. Jahrgangsstufe. Neben einer Wirkungsuntersuchung wurden im Rahmen der Evaluation insbesondere auch die Modellbildungsprozesse der Schüler verfolgt und folgender Fragestellung untersucht:

- Wie wirken sich die von den Lernenden verwendeten Modelle auf das Verstehen grundlegender Sachverhalte zur Holographie aus, und welche Trennschärfe ist zwischen alternativen Modellvorstellungen vorhanden?

Der Untersuchung wurde dabei die Hypothese zugrunde gelegt, dass ein sicherer und problemangemessener Modellgebrauch bei den Schülern nachzuweisen ist, die eine ausgeprägte Trennschärfe zwischen den alternativen Modellvorstellungen zeigen, während Schüler ohne Trennschärfe zwischen den Modellen nicht in der Lage sind, die Holographie konsistent zu erklären.

**2. Untersuchungsmethoden**

Im Rahmen der Evaluation dieser Reihe wurden die Wirkungen des Unterrichts zur Holographie mit Hilfe eines Pre-Post-Testdesigns ermittelt. Neben der Videographie des Unterrichts, einem Wissens- und Motivationstest und Schülerinterviews wurde insbesondere die Anfertigung von Concept Maps durch die Schüler genutzt, um Modellbildungsprozesse zu verfolgen.

Die von den Schülern zu bearbeitenden Concept Maps enthielten zehn vorgegebene und fest platzierte Begriffe, die aus den Ergebnissen einer Vorstudie mit Universitätsstudenten abgeleitet wurden. Den Probanden wurde dabei aufgetragen, die Propositionspfeile direkt nach dem Eintragen zu beschriften. Die Modalmaps von Vor-, Nach- sowie Langzeittest wurden in Anlehnung an [4] aus den Concept Map-Daten der Studenten erstellt, die daraufhin sowohl and the structural parameters of the student maps were compared. In total 54 students participated in this study.

bezüglich Inhalt wie auch Ausprägung ihrer Strukturparameter verglichen wurden. Insgesamt nahmen 54 Schüler an dieser Studie teil.

## 3. Results

## 3. Ergebnisse

In the pre-test, students demonstrated only rudimentary knowledge of holography; yet, the concepts of holograms and inference patters were linked by logically correct propositions. The expressed knowledge was however reproduced from the students' stored knowledge. The participating groups were not familiar with the formation of interference patters or the physical role of reference and object waves. The propositions drawn from the expression '*laser*' confirm that the process of illumination is described with a leaning towards photography. Optical conceptions of rays are linked to this, which is also reflected in the naïve-photographic misconception of numerous students concerning the functions of lenses. Reconstruction is similarly misunderstood.

Im Vortest zeigten die Schüler nur rudimentäre Kenntnisse der Holographie. Zwar wurden die Begriffe Hologramm und Interferenzmuster durch sinnrichtige Propositionen verbunden. Das dadurch ausgedrückte Wissen wurde jedoch lediglich dem Vorratswissen der Schüler reproduzierend entnommen. Kenntnisse über das Zustandekommen des Interferenzmusters und die physikalische Rolle von Referenz- und Objektwelle waren in den Lerngruppen nicht vorhanden. Die vom Laserlicht ausgehenden Propositionen belegen, dass der Beleuchtungsvorgang in Anlehnung an die Photographie beschrieben wird. Damit werden strahlenoptische Vorstellungen verbunden, was sich auch in der naiv-photographischen Fehlvorstellung zahlreicher Schüler bezüglich der Linsenfunktion widerspiegelt. Ebenso unverstanden ist der Rekonstruktionsvorgang.

As the final test shows, the structural development of the modal maps is evidently triggered by the lessons.

Deutlich ist die durch den Unterricht bewirkte strukturelle Weiterentwicklung der Modalmaps zu verfolgen, wie der Nachtest belegt.

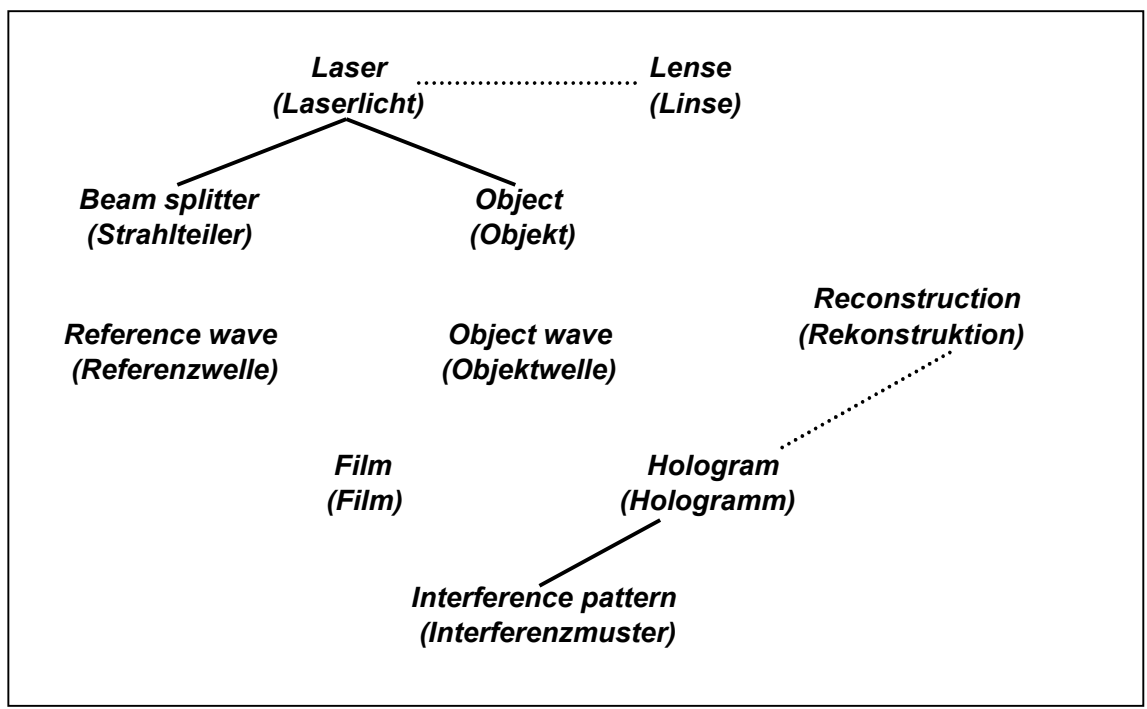

*Figure 1:* Modal map of the pre-test.
*Abbildung 1:* Modalmap des Vortests.
(correct propositions / korrekte Propositionen: ———
false propositions / fehlerhafte Propositionen: ············ )

The isolated partial networks of the pre-test were joined to form an entire network at the final text, which not only had noticeably more propositions, but also a considerably higher number of correct pro-

Die isolierten Teilnetze des Vortests wurden zu einem Gesamtnetz verknüpft, welches nicht nur deutlich mehr Propositionen, sondern auch einen erheblich höheren Anteil an korrekten Propositionen positions. As far as content is concerned, the optical processes of waves that lead to interference patterns were described correctly and the function of the beam splitter and especially the reference wave were described in a physically appropriate manner. The hexagonal structure of the modal map perfectly represents the foundations of holography. However, the participating groups still do show signs of misconceptions as far as reconstruction is concerned. Similarly, the functions of lenses are only understood by a fraction of the participating students, since this topic is still partially described in the technologically problematic context of photography.

aufweist. Inhaltlich wurden die wellenoptischen Prozesse, die zum Interferenzmuster führen, richtig wiedergegeben und die Funktion des Strahlteilers und insbesondere der Referenzwelle physikalisch angemessen beschrieben. Die hexagonale Struktur des Modalmaps bildet die Grundlagen der Holographie vollständig ab. Allerdings zeigen die Lerngruppen noch Fehlvorstellungen, was den Rekonstruktionsvorgang betrifft. Ebenso ist die Linsenfunktion nur einem Teil der Probanden einsichtig, da diese teilweise immer noch fachlich problematisch im Kontext der Photographie beschrieben wird.

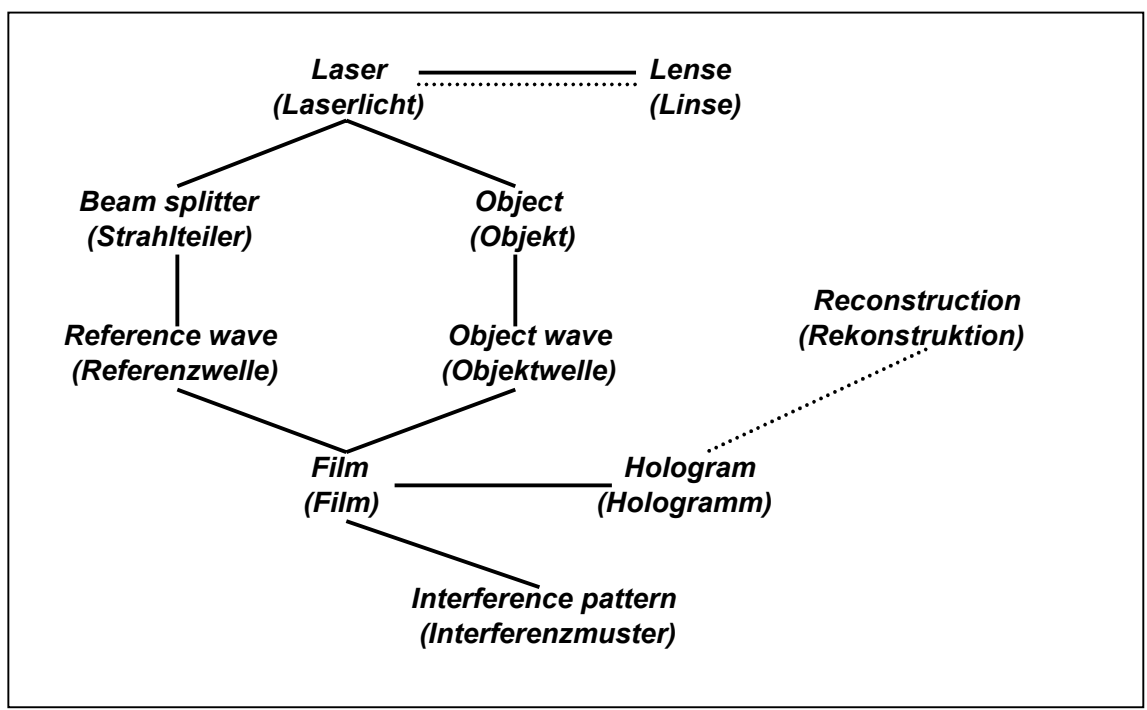

*Figure 2:* Modal map of the final test.
***Abbildung 2:*** *Modalmap des Nachtests.*
*(correct propositions / korrekte Propositionen: ———*
*false propositions / fehlerhafte Propositionen: ············ )*

The reconciliation of data with the other instruments of examination [2] indicates that there was not only a highly significant increase in knowledge, but also that this knowledge was consolidated. The implementation of this knowledge largely depends on whether the students understand the models. This consolidation is also reflected by the data of the structural parameters (see figure 3).

During student interviews it became evident that cognitive conflicts exist between the models, since many participating students used the terminology of the wave model within the context of the ray model. Most of the students correctly pinpointed the experimental features. However, only those students, who had developed a distinct capacity for distinguishing between the models were able to

Der Abgleich der Daten mit den anderen Untersuchungsinstrumenten belegt, dass nicht nur ein höchst signifikanter Zuwachs des Wissens zur Holographie erfolgte, sondern auch eine Festigung des Wissens eintrat. Der Umgang mit diesem Wissen ist jedoch in starkem Maße abhängig vom Modellverständnis, das die Schüler aufweisen. Diese Konsolidierung des Wissens spiegelt sich ebenfalls in den Daten der Strukturparameter (siehe Abbildung 3) wider.

In den Schülerinterviews offenbarten sich zahlreiche Modellkonflikte, da zahlreiche Probanden die Terminologie des Wellenmodells im Kontext des Strahlenmodells nutzten. Experimentelle Gegebenheiten wurden von den meisten Schülern in den Interviews korrekt aufgezeigt. Jedoch konnten bei geforderten Transferleistungen nur diejenigen Schüler sinngerecht mit der Aufgabenstellung umgehen, die eine properly deal with the required task of transfer. These students were able to conduct their arguments in a conscious and problem-oriented manner, to recognize the limits of the models and not only to distinctly designate change in models and in levels of models, but also to apply and substantiate them in a physically logical manner.

Students without this capacity for distinguishing randomly jumped between different model conceptions and were able to impart consistent patterns of explanation less frequently than students who had a more distinct capacity. This was also reflected in the number of correctly marked propositions in their concept maps.

ausgeprägte Trennschärfe zwischen den Modellen entwickelt hatten. Diesen Schülern gelang es, bewusst und problemorientiert zu argumentieren, die Modellgrenzen zu erkennen und Wechsel von Modellen und Modellebenen nicht nur ausdrücklich zu benennen, sondern diese physikalisch sinnrichtig anzuwenden und zu begründen.

Schüler ohne diese Trennschärfe sprangen beliebig zwischen unterschiedlichen Modellvorstellungen hin und her und konnten ein konsistentes Erklärungsmuster weit seltener liefern als Schüler mit ausgeprägter Trennschärfe. Dies spiegelte sich auch in der Anzahl der von ihnen korrekt beschrifteten Propositionen ihrer Concept Maps wider.

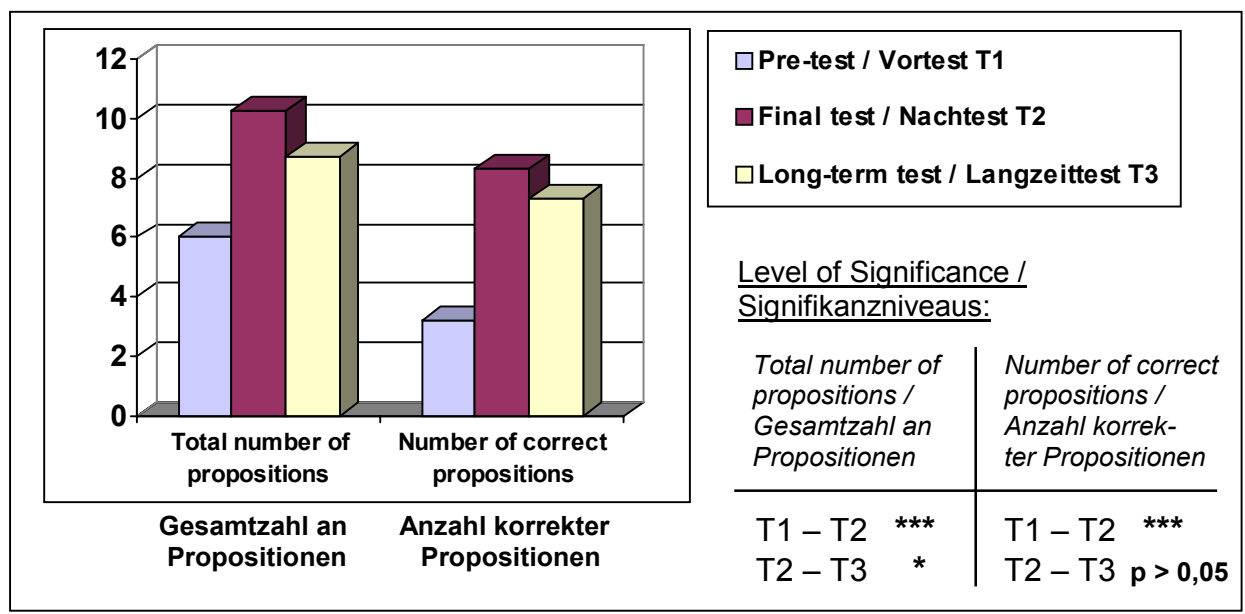

*Figure 3:* The development of the numbers of propositions.
*Abbildung 3:* Zeitliche Entwicklung der Anzahl an Propositionen.

### 4. Conclusions and implications

The hypotheses was ultimately confirmed. The capacity demonstrated by the students for distinguishing between the models of light (e.g. ray model, wave model, classic and quantum-optical particle model as well as the phasor model) is of great importance.

Therefore, more attention should be paid to the development of a meta-conceptual awareness in physics classes, so that the explanations of diverse models and model levels will be examined and discussed. As the detailed studies of our research group show, it is of advantage that these model discussions and reflections are deliberated already in the beginner's lessons of physics and that the students have the opportunity to scrutinize the limits of models already as they are introduced.

### 4. Schlussfolgerungen und Konsequenzen

Die Hypothese konnte bestätigt werden. Der von den Probanden gezeigten Trennschärfe zwischen den Modellen des Lichts (Strahlenmodell, Wellenmodell, klassisches und quantenoptisches Teilchenmodell sowie Zeigermodell) kommt eine entscheidende Bedeutung zu.

Es sollte deshalb im Physikunterricht weit stärker auf die Ausbildung eines metakonzeptuellen Bewusstseins geachtet werden, so dass Erklärungsmuster unterschiedlicher Modelle problematisiert und unterschiedliche Modellebenen angesprochen werden. Wie weitergehende Untersuchungen unserer Arbeitsgruppe zeigen, ist es von Vorteil, wenn diese Modelldiskussionen und Modellreflektionen bereits im Anfängerunterricht Physik thematisiert werden und Schüler Modellgrenzen bereits mit Einführung von Modellen hinterfragen.